%% file: ms_mnras.tex
\documentclass[useAMS,usenatbib]{mn2e}
\usepackage{epsfig,bm}

\newcommand{\be}{\begin{eqnarray}}
\newcommand{\ee}{\end{eqnarray}}
\newcommand{\beq}{\begin{equation}}
\newcommand{\eeq}{\end{equation}}
\def\simless{\mathbin{\lower 3pt\hbox
      {$\rlap{\raise 5pt\hbox{$\char'074$}}\mathchar"7218$}}}
\def\simgreat{\mathbin{\lower 3pt\hbox
      {$\rlap{\raise 5pt\hbox{$\char'076$}}\mathchar"7218$}}} 

\renewcommand{\vec}[1]{\mbox{\boldmath $\displaystyle #1$}}

\newcommand{\grad}{{\mbox{\boldmath $\nabla$}}}

\newcommand{\apj}{ApJ}
\newcommand{\apjl}{ApJL}

\newcommand{\mnras}{MNRAS}

\newcommand{\pasp}{PASP}

\newcommand{\aap}{A\&A}

\newcommand{\procspie}{Proceedings of the SPIE}

\title{ The Radial Velocity Signature of Tides Raised in Stars Hosting Exoplanets }

\author[ P.~Arras, J.~Burkart, E.~Quataert, N.N.~Weinberg ]{
\parbox[t]{\textwidth}{ 
Phil Arras$^{1}$\thanks{E-mail: arras@virginia.edu (PA); burkart@berkeley.edu (JB);
eliot@astro.berkeley.edu (EQ); nevin@mit.edu (NNW)}, Joshua Burkart$^{2,3}$,
Eliot Quataert$^{2,3}$ and Nevin N.~ Weinberg$^{4}$ }
\vspace*{6pt} \\
$^{1}$Department of Astronomy, University of Virginia,
P.O. Box 400325, Charlottesville, VA 22904-4325, USA \\
$^2$Department of Physics, 366 LeConte Hall, University of California,
Berkeley, CA 94720, USA \\
$^3$Department of Astronomy \& Theoretical Astrophysics
  Center, 601 Campbell Hall, University of California Berkeley, CA
  94720, USA \\
$^4$Department of Physics and MIT Kavli Institute, MIT, 77 Massachusetts Avenue,
Cambridge, Massachusetts 02139, USA 
}

\begin{document}

\date{}

\pagerange{\pageref{firstpage}--\pageref{lastpage}} \pubyear{2011}

\maketitle

\label{firstpage}

\begin{abstract}

Close-in, massive exoplanets raise significant tides in their stellar
hosts. We compute the radial velocity (RV) signal due to this fluid
motion in the equilibrium tide approximation. The predicted radial
velocities in the observed sample of exoplanets exceed $\rm 1\ m/s$
for 17 systems, with the largest predicted signal being $\rm \sim
30\ m\ s^{-1}$ for WASP-18 b. Tidally-induced RV's are thus detectable
with present methods.
Both tidal fluid flow and the epicyclic motion of a slightly eccentric orbit produce an RV
signal at twice the orbital frequency. If care is not taken, the tidally induced RV may, in some
cases, be confused with a finite orbital eccentricity. Indeed, WASP-18 b is reported to have an
eccentric orbit with small $e=0.009$ and pericenter longitude $\omega=-\pi/2$. Whereas 
such a close alignment of the orbit and line of sight to the observer
requires fine tuning, this phase in the RV signal is naturally
explained by the tidal velocity signature of an $e=0$ orbit.
Additionally, the equilibrium tide estimate for the amplitude is
in rough agreement with the data. Thus the reported eccentricity
for WASP-18 b is instead likely a signature of the tidally-induced
RV in the stellar host. Measurement of both the orbital and
tidal velocity for non-transiting planets may allow planet mass and
inclination to be separately determined solely from radial velocity
data.  We suggest that high precision fitting of RV data should
include the tidal velocity signal in those cases where it may affect
the determination of orbital parameters.

\end{abstract}



\section{Introduction}

Increasing RV precision allows the detection of smaller
and/or more distant planets. The current state of the art is Doppler
measurement errors in the range $ \sim \rm 1 \ m\ s^{-1}$
(e.g. \citealt{1996PASP..108..500B,2006SPIE.6269E..23L}). At
this level of measurement precision, other physical effects
become potentially detectable besides the Keplerian
stellar orbital motion: the Rossiter-McLaughlin effect
(e.g. \citealt{2011EPJWC..1105002W}), additional planets perturbing the
orbit (e.g. \citealt{2001ApJ...551L.109L}), ``jitter" due to convective
overturn motions in the stellar envelope \citep{2005PASP..117..657W} and
solar-like acoustic waves \citep{2007CoAst.150..106B}.  In this paper
we investigate an additional effect, the time-dependent spectroscopic
shift due to fluid motions in the star forced by the planetary tide.

Previous investigations have highlighted that tides raised in the
star by the planet may give observable ellipsoidal (flux) variation
\citep{2003ApJ...592.1217S, 2003ApJ...588L.117L,2008ApJ...679..783P}.
NASA's Kepler Mission has recently announced the first detection of
ellipsoidal variation due to a planet \citep{2010ApJ...713L.145W}.

The tidal velocity signal has been discussed by
\citet{2002A&A...384..441W} for the case of a massive star primary and
stellar mass compact object secondary. \citet{1977AcA....27..203D} gives
formulae to convert the surface fluid motions to disk averaged radial
velocities along the line of sight to the observer. In this paper we
discuss the tidally induced fluid motions due to a planetary companion
with short orbital period, for which the tidal velocity may be observable.

The tidal velocity is computed in \S \ref{sec:rv}. Distinguishing between
a small orbital eccentricity and the tidal velocity is discussed in
\S \ref{sec:distinguish}. Equilibrium tide estimates for the tidal
velocity of known exoplanets, and the cases of WASP-18 b and WASP-33 b,
are discussed in \S \ref{sec:discussion}.


\section{ tidal and orbital radial velocities }
\label{sec:rv}

Consider a planet of mass $M_p$ in a Keplerian orbit around a star
of mass $M$ and radius $R$. The semi-major axis, eccentricity, orbital
frequency, orbital separation and true anomaly are denoted $a$, $e$, $n=(
G(M+M_p)/a^3 )^{1/2}=2\pi/P_{\rm orb}$, $d$ and $f$, respectively. The
orbital separation is given by $d=a(1-e^2)/(1+e\cos f)$, where $f=0$
at pericenter. We consider the star to be non-rotating. Corrections due to stellar
rotation will be discussed in \S \ref{sec:discussion}.

A rough estimate of the tidal velocity may be obtained as
follows. The ratio of the tidal force to the internal gravitational
force gives a dimensionless strength of the tide: $\epsilon  \equiv
(M_p/M)(R/d)^3$.
The height of the tide is then $\epsilon R$. Distant observers
see the fluid oscillate at harmonics of the orbital frequency, $n$.
The tidal velocity is then roughly $v_{\rm tide}  \simeq  \epsilon
n R \sim \rm (1-10)\ m\ s^{-1}$ for massive, close-in planets (see
eq.~[\ref{eq:vtidecirc}]).

\begin{figure}
\begin{center}
\epsfig{file=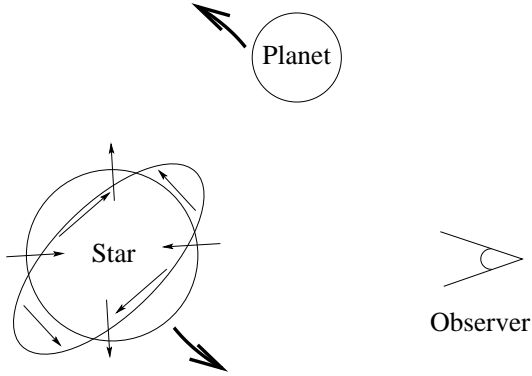,height=5.0cm}
\end{center}
\caption{ Diagram showing the orientation of the orbital and fluid motion,
as viewed looking down on the orbital plane.
The planet and unperturbed star are shown as circles, with boldface arrows in the
direction of the orbital motion about the center of mass. The ellipsoid
represents the equilibrium tide deformation of the star, the long axis of
which follows the planet in its orbit. The thin arrows show the direction
of vertical and longitudinal fluid motion at intervals of $\pi/4$.
}
\label{fig:eqtide}
\end{figure}

The tidal bulge is commonly pictured as an ellipsoid rotating at
the angular velocity of the binary. This is a good approximation
if the star's rotation is synchronized to the orbit.
However, exoplanet
host stars typically rotate much slower than the orbital frequency
of the planet,
causing fluid elements to experience a time-changing tidal force.
This force induces both vertical and horizontal fluid motions which
cannot be modeled as rigid rotation, as we now discuss in more
detail.

Figure \ref{fig:eqtide} shows the relative signs of the tidal flow and
orbital velocity in the simplest case of a circular orbit.
The circle represents the unperturbed star, and the ellipsoid
represents the tidally deformed star, with the deformation following the
planet in its orbit. As the bump approaches a particular point on the
surface, the fluid motion is upward. After the bump passes the fluid
moves back down. In the lowest-order case of a quadrupole, there are
two bumps, and so the fluid goes up and down twice per orbit. Starting
the planet in between star and observer, the star's orbital and tidal
velocity are zero. Advancing the orbit slightly, the star is moving toward
the observer, while the tidal velocity is away from the observer. Other
phases can be deduced similarly. For a circular orbit, only the relative
angle between observer and planet as seen by the star matters for both
$v_{\rm orb}$ and $v_{\rm tide}$.  However, including finite orbital
eccentricity introduces the pericenter longitude into the problem,
meaning that the second and higher harmonic components of $v_{\rm orb}$
(present only for $e\neq 0$) have a more complicated angular dependence.

In order to distinguish the tidal RV from the epicyclic motion of an
eccentric orbit, it is necessary to understand the orbit's orientation
and the various sign conventions for RV.  Define a coordinate system
with the origin at the center of the star, with the planet orbiting in the
x-y plane with separation vector from star to planet given by $\vec{d}=
d\left( \vec{e}_x\,\cos f + \vec{e}_y\,\sin f \right) $. To describe the
fluid motion in the star, spherical coordinates $(r,\theta,\phi)$ are
used for the position $\vec{x}$. The coordinates of the planet's orbit are
then $(d,\pi/2,f)$. The direction to the observer as seen from the star,
$\vec{n}_o$, can be specified in two ways.  The calculation of the tidal
signal is most convenient using our polar coordinate system centered
on the star, in which $\vec{n}_o = \sin\theta_o (\vec{e}_x \cos\phi_o
+ \vec{e}_y \sin \phi_o) + \vec{e}_z \cos\theta_o$.  However, the RV
measurement involves the Euler angles for inclination $i$ and longitude of
pericenter $\omega$ as seen by the observer. To relate $(\theta_o,\phi_o)$
and $(i,\omega)$ we use the transformation matrix from the orbit frame,
$(x,y,z)$, to the observer frame $(X,Y,Z)$ \citep{1999ssd..book.....M}.
Choosing the observer to lie along the $+Z$ direction, and $X-Y$ to be
the plane of the sky, the observer direction as seen from the star is
$\vec{n}_o  = \sin i (\vec{e}_x\ \sin \omega + \vec{e}_y\ \cos \omega)
+ \vec{e}_z\ \cos i$.  Equating the two expressions for $\vec{n}_o$,
we find $\theta_o = i$ and $\phi_o = \pi/2 - \omega$.

We follow the conventions that the Doppler shift velocity is positive for motion away
from the observer, and that the separation
vector points from star to planet.
The vector from the center of mass to the star is then
\be
\vec{x}_\star & = & - \left( \frac{M_p}{M_p+M} \right)\ d\ \left( \vec{e}_x \cos f + \vec{e}_y
\sin f \right).
\ee
The RV of the star due to its orbital motion becomes
\be
v_{\rm orb}(t) & = & - \vec{n}_o \cdot \dot{\vec{x}}_\star
\nonumber \\ & = &
- K_{\rm orb}
\left( \sin \left( f -\phi_o \right) - e\sin\phi_o \right)
\\ & = &
K_{\rm orb}
\left( \cos \left( f + \omega \right) + e\cos \omega \right),
\label{eq:vorb}
\ee
where
\be
K_{\rm orb} & = & \left( \frac{M_p}{M_p+M} \right)\
\left( \frac{n a \sin i}{\sqrt{1-e^2}} \right)
\ee
is the semi-amplitude of the orbital velocity.  Equation (\ref{eq:vorb}) is the
standard formula used to fit RV data \citep{2010exop.book...15M}.

Next we turn to the tidal velocity. The disk-averaged RV away from the observer is defined to
be \citep{1977AcA....27..203D}
\be
v_{\rm tide} & = & \frac{ \int d\Omega\ h( \vec{n} \cdot \vec{n}_o )\
\vec{n} \cdot \vec{n}_o\  \left( - \dot{\vec{\xi}} \cdot \vec{n}_o \right) }
{\int d\Omega\ h( \vec{n} \cdot \vec{n}_o )\
\vec{n} \cdot \vec{n}_o },
\label{eq:avgvpar}
\ee
where the integral is over the unperturbed surface of the star.
Here $\vec{\xi}$ and $\vec{\dot{\xi}}$ are the displacement vector and
velocity of the fluid relative to the background star. The
limb darkening function, $h(\mu)$, is normalized as $\int_0^1 d\mu
\mu h(\mu)=1$, making the denominator $2\pi$.  Since $\dot{\vec{\xi}}
\cdot \vec{n}_o$ is the velocity toward the observer, the negative sign
in equation (\ref{eq:avgvpar}) gives the velocity away from the observer.

The tidal potential $U(\vec{x},t)$ and fluid displacement can be expanded in spherical harmonics as
\be
U(\vec{x},t)
& = &  - GM_p \sum_{\ell \geq 2} \sum_{m=-\ell}^\ell
\frac{4\pi}{2\ell+1} \frac{r^\ell}{d^{\ell+1}}
  \nonumber \\ & \times &
Y^*_{\ell m}(\pi/2,f)Y_{lm}(\theta,\phi),
\label{eq:Ulm}
\\
\xi_r(\vec{x},t)  & = &   \sum_{\ell m} \xi_{r,\ell m}(r,t) Y_{\ell m}(\theta,\phi)
\\
\xi_\theta(\vec{x},t)  & = &   \sum_{\ell m} \xi_{h,\ell m}(r,t)
\frac{\partial Y_{\ell m}(\theta,\phi)}{\partial \theta}
\\
\xi_\phi(\vec{x},t)  & = &   \sum_{\ell m} \xi_{h,\ell m}(r,t)
\frac{1}{\sin\theta}\frac{\partial Y_{\ell m}(\theta,\phi)}{\partial \phi},
\ee
where $\xi_{r,\ell m}$ and $\xi_{h,\ell m}$ are the spherical
harmonic coefficients of the vertical and horizontal motions, found
as the forced response to $U$.

\citet{1977AcA....27..203D} shows that the integral in equation (\ref{eq:avgvpar})
may be computed by using rotation matrices to express the
spherical harmonics in the orbit frame in terms of another coordinate
system oriented toward the observer. The result can be written in the form
\be
v_{\rm tide} & = & - \sum_{\ell \geq 2,m}
\left( u_\ell \dot{\xi}_{r,\ell m}(t) + v_\ell \dot{\xi}_{h,\ell m}(t) \right)
Y_{\ell m}(\theta_o,\phi_o),
\label{eq:rv1}
\ee
where the limb-darkening integrals are
\be
u_\ell & =&  \int_0^1 d\mu \mu^2 P_\ell(\mu) h(\mu)
\label{eq:ul}
\\
v_\ell & =&  \int_0^1 d\mu \mu (1-\mu^2) \frac{dP_\ell(\mu)}{d\mu} h(\mu),
\label{eq:vl}
\ee
and $P_\ell(\mu)$ is a Legendre polynomial.
For Eddington limb darkening, $h=1+3\mu/2$, \citet{1977AcA....27..203D}
gives a table of values including $u_2=0.321$, $v_2=0.775$, $u_3=0.127$
and $v_3=0.593$. We found $u_\ell$ and $v_\ell$ to vary weakly with
changes in the limb darkening law.

For the fluid motion, we employ the {\it equilibrium tide approximation},
in which the fluid motion is incompressible and follows gravitational
equipotentials (e.g.\ \citealt{1989ApJ...342.1079G}).\footnote{ We ignore
the small contribution to the gravitational potential arising from density
perturbations within the star (the Cowling approximation).}
For the fluid to follow the combined equipotential
of the background star and tidal potential, the vertical displacement
must be
\be
\xi_{r,eq}(\vec{x},t) & = & - \frac{U(\vec{x},t)}{g},
\label{eq:xireq}
\ee
where $g(r)=Gm(r)/r^2$ is the gravity of the background model.  The components
of the horizontal displacement are then determined by the condition
$\grad \cdot \vec{\xi}=0$:
\be
\xi_{h,\ell m} & = & \frac{1}{\ell(\ell+1)r} \frac{d}{dr} \left( r^2 \xi_{r,eq,\ell m} \right)
\nonumber \\ & \simeq & 
\left( \frac{\ell+4}{\ell(\ell+1)} \right) \xi_{r,eq,\ell m}
\label{eq:xiheq}
\ee
where in the second expression we have used $\xi_{r,eq,\ell m} \propto
r^{\ell+2}$ near the surface, where the interior mass is nearly constant.

Plugging equations (\ref{eq:xireq}), (\ref{eq:xiheq}) and (\ref{eq:Ulm})
into equation (\ref{eq:rv1}),
using the spherical harmonic addition formula to sum over $m$,
and taking the time derivative
gives the general formula for tidal RV under the equilibrium
tide approximation:
\be
&& v_{\rm tide}  =  R \frac{M_p}{M} \sum_{\ell=2}^\infty f_\ell \left( \frac{R}{d} \right)^{\ell+1}
\left[ \left( \ell+1 \right) P_{\ell}\left(\cos\gamma \right)
\frac{\dot{d}}{d}
\nonumber \right. \\ & &\qquad\qquad \left.
+ \sin \theta_o \sin(f-\phi_o) \frac{dP_\ell(\cos\gamma)}{d(\cos\gamma)} \dot{f} \right].
\label{eq:vtide}
\ee
Here $\cos\gamma = \sin\theta_o \cos(f-\phi_o)$
is the angle between planet and observer as seen by the star. The parameter 
$f_\ell = \left( u_\ell \dot{\xi}_{r,\ell m}(t) + v_\ell \dot{\xi}_{h,\ell m}(t) \right)/
\dot{\xi}_{r,eq,\ell m}$
contains information about limb darkening and the size of the surface fluid motions,
and takes on the value $f_\ell = u_\ell + v_\ell (\ell+4)/[\ell(\ell+1)]$ in the 
equilibrium tide approximation. This parameter is the main uncertainty in our model.
In a more exact treatment of the fluid motions, $f_\ell$ would be dependent on the 
stellar structure, the orbital period, and the stellar rotation rate.

Equation (\ref{eq:vtide}) can be simplified considerably for a circular orbit keeping only the
dominant $\ell=2$ term:
\be
&& v_{\rm tide} = \frac{3}{2} nR\ \frac{M_p}{M}\ \left( \frac{R}{a} \right)^3\
f_2
\sin^2 \theta_o \sin \left[ 2(n t - \phi_o) \right]
\\
& \simeq & 1.13\ {\rm m\ s^{-1}}\ \left( \frac{M_p}{M_J} \right) \left( \frac{M_\odot^2}{M(M+M_p)}
\right)
\nonumber \\ & \times &
\left( \frac{R}{R_\odot}\right)^4 \left( \frac{1\ \rm day}{P_{\rm orb}} \right)^3
\sin^2 \theta_o \sin \left[ 2(nt - \phi_o) \right],
\label{eq:vtidecirc}
\ee
where in the second step we used Eddington limb darkening for which
$f_2\simeq 1.10$.  Equation (\ref{eq:vtidecirc}) shows that, not surprisingly,
the tidal velocity signal is largest for massive planets in short period
orbits around stars with large radii.

We end this section with a brief discussion of our choice to use the
equilibrium tide approximation. The derivation of equations (\ref{eq:xireq})
and (\ref{eq:xiheq}) from the fluid
equations proceeds first by ignoring inertia and setting the forcing
frequency to zero (e.g.\ \citealt{1989ApJ...342.1079G}), and second by solving the equations in
a radiative zone, where the Brunt-Vaisalla frequency $N^2 > 0$. In a
convection zone, it's well known (e.g.\ \citealt{1998ApJ...502..788T})
that the same derivation does not hold if one sets $N^2 = 0$,
and deviations from the equilibrium tide are expected, even for
low frequencies. However, $\xi_r \simeq \xi_{r,eq}$ is still a good
approximation at the surface, since the surface of the star must be on
an equipotential if one ignores fluid inertia. The same cannot be said
of $\xi_h$. The value of $\xi_h$ must adjust to cause $\xi_r \simeq
\xi_{r,eq}$ at the surface and also the base of the convection zone,
and in general $\xi_h \neq \xi_{h,eq}$ exactly.

To understand the behavior of $\xi_h$ in more detail we performed a
number of integrations of the equations for adiabatic fluid perturbations
\citep{1989nos..book.....U} forced by the gravitational tide, for stars in
the mass range $M=1.0-1.4 M_\odot$. We find that $\xi_h \sim \xi_{h,eq}$
at roughly the factor of 2 level away from g-mode resonances, with $\xi_h$
showing more variation with tidal frequency than $\xi_r$. More variation with 
frequency occurs with increasing $M$, as the convection zone shrinks,
and g-modes have a more pronounced effect at the surface of the star. Overall,
this implies that the equilibrium tide estimate of $v_{\rm tide}$ is 
typically accurate to a factor of $\simeq 2$, but that the tidally
induced RV can be significantly larger near g-mode resonances.
We use the equilibrium tide approximation here largely for its simplicity,
with the intent of performing more rigorous calculations in the future.


\section{ amplitude and phase }
\label{sec:distinguish}

In order to measure the tidal RV signal, it must be distinct from that of the orbital
RV so as not to confuse the two.
Equations (\ref{eq:vtide}) and (\ref{eq:vorb}) for the tidal and orbital
velocities give the amplitude and phase of all harmonics of the RV
signal for eccentric orbits. At large eccentricity, 
the two signals are clearly distinct, due to the strong tidal RV
dependence on orbital separation.
Moreover, for an eccentric orbit, equation (\ref{eq:vtide}) gives nonzero values even
for a face on orientation. 
The limit of small but finite
eccentricity is more subtle, as we now discuss.

First consider a circular orbit with planet phase angle $\phi=nt$ and
observer position $\phi_o$. As the orbit is circular, the pericenter
longitude $\omega$ is undefined, and the longitudes $\phi$ and $\phi_o$
can be taken with respect to any reference line. We define $t=0$ to occur
when the planet is along the reference line. Inferior conjunction
(planet between star and observer) occurs at $\phi=\phi_o$, and superior
conjunction (planet on opposite side of star) occurs at $\phi=\phi_o +
\pi$. The orbital velocity
\be
v^{(k=1)}_{\rm orb}(t) & = & -  K_{\rm orb} \sin \left( \phi - \phi_o \right)
\label{eq:vorb1}
\ee
is at the first harmonic of the orbital frequency, denoted $k=1$, while 
the tidal velocity is at the second harmonic ($k=2$)
\be
v_{\rm tide}(t) & = & K_{\rm tide} \sin \left[ 2(\phi - \phi_o) \right]
\label{eq:vtide2}
\ee
with amplitude
\be
K_{\rm tide}  & = & \frac{3}{2} nR \frac{M_p}{M} \left( \frac{R}{a} \right)^3
f_2 \sin^2 i.
\ee
In the circular orbit case, the orbital and tidal signals are determined
solely by the angular separation $\phi-\phi_o$.
The signs of the RV in equations (\ref{eq:vorb1}) and (\ref{eq:vtide2})
have been discussed in \S \ref{sec:rv} and Figure \ref{fig:eqtide}.
Just after inferior conjunction, $v_{\rm tide} > 0$ (away from observer)
 and $v^{(k=1)}_{\rm orb} < 0$ (toward observer), while just after
superior conjunction $v_{\rm tide} > 0$ and $v^{(k=1)}_{\rm orb} > 0$.
The combined radial velocity curve for the circular orbit case is
given by equations (\ref{eq:vorb1}) and (\ref{eq:vtide2}), with $\phi=nt$, to be
\be
v^{(k=1)}_{\rm orb}(t) + v_{\rm tide}(t) =
& - &   K_{\rm orb} \sin \left( nt - \phi_o \right)
\nonumber \\ &  + & 
 K_{\rm tide} \sin \left[ 2(nt - \phi_o) \right].
\label{eq:combined}
\ee

Next we determine under what conditions a slightly eccentric orbit, ignoring tides, can mimic
the tidal velocity plus circular orbit velocity in equation (\ref{eq:combined}).
To this end,
equation (\ref{eq:vorb}) can be expanded to ${\cal O}(e)$ as
\be
v_{\rm orb}(t) & \simeq & -  K_{\rm orb} \left[ \sin \left( nt - \phi_o \right)
+ e \sin \left( 2nt - \phi_o \right) \right]
\nonumber \\ & \equiv & 
v^{(k=1)}_{\rm orb}(t) + v^{(k=2)}_{\rm orb}(t).
\label{eq:vorb12}
\ee
Comparing equations (\ref{eq:combined}) and (\ref{eq:vtide2}), the tidal
and epicyclic velocities are both at the
second harmonic ($k=2$) of the orbital frequency. However, they do
not in general have the same amplitude or phase.
In particular, equation (\ref{eq:vorb12}) shows that the $k=2$ orbital
term has a different phase, $nt-\phi_o/2=(nt-\phi_o) + \phi_o/2$,
which is not just the planet-observer angle---there is an additional dependence
on the viewing angle $\phi_o = \pi/2-\omega$.

Furthermore, it is possible for the $k=2$ orbital velocity to mimic the $k=2$ tidal velocity: for the specific choice
\be
K_{\rm tide} &  = & e K_{\rm orb}
\label{eq:ampcond}
\\
\phi_o  =  \pi\quad &\leftrightarrow\ & \quad \omega = - \pi/2,
\label{eq:phasecond}
\ee
the tidal and $k=2$ orbital signals are the same. That is, the tidal velocity
is degenerate with an eccentric orbit (of appropriate $e$; eq.~[\ref{eq:ampcond}])
in which the long axis of the orbit is along the line of sight to the observer.

Given that an eccentric orbit can mimic the tidal RV
signal, we suggest that when the values of $e K_{\rm orb}$
and $\omega$ inferred from fitting solely to an orbital RV
model roughly satisfy equations (\ref{eq:ampcond}) and (\ref{eq:phasecond}),
it is reasonable to presume that what is being detected is the tidal
velocity, not a finite eccentricity.
In that case, the value of $e$ from the fit does not describe the orbit,
but rather gives the magnitude of the tidal velocity 
$K_{\rm tide} = e K_{\rm orb}$.
 Since the orbit can a priori have
any orientation $0 \leq \omega \leq 2\pi$, 
fine tuning is required in order to obtain the value $\omega = - \pi/2$ implied
by the finite eccentricity interpretation of the RV data. By contrast, the 
tidal velocity of an $e=0$ orbit naturally explains the second harmonic
of the RV data.



\section{Discussion and Conclusions}
\label{sec:discussion}

\input{tab1_mnras.tex}

To evaluate the magnitude of the tidally-induced radial velocity
(eq.~[\ref{eq:vtide}]) for observed exoplanets, data were taken from
the Extrasolar Planet Encyclopedia (http://exoplanet.eu/catalog.php)
with two exceptions. Updated values for WASP-18 b were taking from
\citet{2010A&A...524A..25T}, and for WASP-33 b we include the
eccentricity fit from \citet{2011MNRAS.tmp.1075S}, and use the upper limit
on the mass as an estimate of the mass itself. For transiting
planets, the measured inclination is used. For non-transiting planets,
we set $\theta_o=\pi/2$ and use the measured $M_p\sin i$ in place of
$M_p$.  Parameters used and the tidal velocity semi-amplitude, $K_{\rm
  tide}$, found by evaluating equation (\ref{eq:vtide}) over an orbit,
are presented in Table \ref{tab:estimates} for planets with $K_{\rm
  tide}>1\ {\rm m\ s^{-1}}$. For the eccentric orbit cases, the $k=2$
orbital velocity amplitude $e K_{\rm orb}$ is given to compare to
$K_{\rm tide}$.  Most cases show massive, close-in planets in nearly
circular orbits around stars with radii slightly larger than the
Sun. Some planets, such as HAT-P-2 b, XO-3 b and HIP 13044 b have
enhanced signal near pericenter due to large eccentricity.




Table \ref{tab:estimates} shows that the tidal velocity is potentially
detectable in up to 17 known exoplanets for an RV precision of $1\
{\rm m\ s^{-1}}$.  Inspection of the RV curves in the NsTED database
(http://nsted.ipac.caltech.edu/) shows that in many cases, the
precision of the measurements taken was not high enough to allow
measurement of signals in the range $v_{\rm tide}=1-30\ {\rm m\
  s^{-1}}$. Should such measurements become available in the future,
measurement of the tidal velocity will allow a non-trivial check on
the planet mass, inclination, and stellar mass and radius.
For non-transiting planets detected only through RV data,
the tidal signal and orbital motion will allow $M_p$ and $i$ to be
separately determined solely from RV data, since they have different
dependence on inclination. As signal to noise can be built up over
many orbits, follow-up observations on planets already detected may
allow one to measure planet masses for a number of RV and transiting
planets.

We suggest that a model of the tidal RV should be included in high
precision analyses of RV data for close-in planets such as those in
Table \ref{tab:estimates}. At present, the parameter $f_2$ in the
tidal RV (eq.~[\ref{eq:vtidecirc}]) -- which characterizes the properties of the tide
raised by the planet -- could be included as a parameter in the
fit. In the future, more detailed calculations of the tidal response
of the star to its companion may provide accurate values of $f_2$ for
the required range of stellar masses and orbital periods.  We have for
simplicity focused on the equilibrium tide predictions ($f_2 \simeq
1.1$) in this paper.

WASP-18 b clearly stands out as the best candidate for an existing
detection of the tidal velocity, as the claimed orientation of the
eccentric orbit, $\omega\simeq-\pi/2$ \citep{2010A&A...524A..25T}, is in
excellent agreement with the phase predicted by the tidally-induced
RV.  Moreover, the equilibrium tide prediction for the
amplitude is within a factor of 2 of the inferred $e K_{\rm orb}$.
Quantitatively, we find that instead of the equilibrium tide value,
$f_2^{\rm (eq\ tide)}=1.1$, the value measured from the data for
WASP-18 b is $f_2^{\rm (data)} = 1.1 \times (15.38/31.94) = 0.53$.  In
our preliminary numerical calculations of the full tidal response of a
star, we have found that the equilibrium tide approximation is only
accurate at the factor of 2 level for calculating the tidal RV signal
(particularly for the horizontal motion at the surface).  Thus we
regard the agreement between the data and the equilibrium tide
prediction of the RV amplitude for WASP-18 b as satisfactory.  Given
this agreement, we argue that the simplest interpretation of the data
is that the signal detected in WASP-18 b is the tidal velocity, rather
than a finite orbital eccentricity, and that the true orbital
eccentricity has a value $e \ll 0.009$.  Due to the large value of
$R/a$, higher harmonics of the orbital frequency due to
multipole orders $\ell=3,4$ may be detectable in the
data as well. 



WASP-33 b is another candidate in which tidal RV may have been
mistaken for an eccentric orbit. \citet{2011MNRAS.tmp.1075S} initially
fitted an eccentric orbit, finding $e=0.174$ and $\omega = -
89^\circ$. While the value of $\omega$ is strongly suggestive of tidal
velocity, the large amplitude would require fluid motions much larger
than the equilibrium tide (see Table \ref{tab:estimates}). While
larger amplitude tidal flow is expected for higher mass stars, due to
their thin surface convection zones, this factor of $\simeq 20$
increase in amplitude likely requires a near resonance with a g-mode.

The case of WASP-33 b brings up the issue of how stellar rotation affects the tidal
RV signal. While most planet-host stars have slow, sub-synchronous rotation
(even WASP-18 b, an A star) WASP-33 b is an example of an A star that rotates
more rapidly than the orbit. Stellar rotation can be included in the analysis
in two places, through alteration of the fluid motions, mainly through
the Coriolis force, and also by changing the expression for the fluid motion
at the perturbed surface of the star. As the equilibrium tide ignores fluid
inertia, the equilibrium tide displacements are unchanged for a rotating star.
The second way in which rotation changes the radial velocity is that 
rather than using $\dot{\vec{\xi}}$
in equation (\ref{eq:avgvpar}), as is appropriate
for a non-rotating star, the Lagrangian velocity perturbation
\be
\Delta \vec{v} & = & \dot{\vec{\xi}} + \vec{v} \cdot \grad 
\vec{\xi}
\label{eq:ldv1}
\\ & = & 
\vec{e}_i \left( \frac{\partial }{\partial t} + \Omega \frac{\partial }{\partial \phi}
\right)\xi_i + \vec{\Omega} \times \vec{\xi}
\label{eq:ldv2}
\ee
must be used, which includes the effect of (here uniform) rotation ($\vec{v}
= \vec{\Omega} \times \vec{x}$) in the background star. The bracketed term 
in equation (\ref{eq:ldv2}) is the time derivative in the frame co-rotating with the star.
For a star with rotation synchronized to the orbit, this term is zero as the tide
is time-independent in the fluid frame. The second
term in eq.\ref{eq:ldv2} represents the rigid rotation of the deformed surface, and
is present even for a synchronized star.

We have evaluated the form of equation (\ref{eq:ldv2})
for the case of a circular orbit with rotation
and orbital angular momentum axes aligned. The resulting expression shows that
$n \rightarrow n-\Omega$ in the time derivative term, and new terms $\propto \Omega$
arise from the piece $\vec{\Omega} \times \vec{\xi}$. The expected phase of the radial
velocity signal should still be $\omega = - \pi/2$ for the $\Omega < n$ case. However,
it may be possible for the tidal RV phase to shift by $180^\circ$ when $\Omega > n$. 
A more detailed calculation is necessary to settle this issue.
For WASP-18 b, rotation may introduce a small amplitude correction
($\sim 15\%$). The amplitude correction for WASP-33 b is expected to be order unity.
We plan to include rotation in a future study.


\section{Acknowledgments}

We thank Joergen Christensen-Dalsgaard, Geoff Marcy, Amauri Triaud and
Jeff Valenti for useful discussions.  This work was supported by NSF
AST- 0908873 and NASA NNX09AF98G. PA is an Alfred P.  Sloan Fellow, and
received support from the Fund for Excellence in Science and Technology
from the University of Virginia. EQ was supported in part by
the David and Lucile Packard Foundation. JB is an NSF Graduate Research Fellow.


\label{lastpage}

\end{document}

%% file: tab1_mnras.tex
\begin{table*}
\caption{ Estimates of Tidal Velocity }
\label{tab:estimates} 
\begin{tabular}{lccccccccc}
\hline
planet name &  $M_p[\sin i]/M_J$ & $P_{\rm orb}/\rm day$ & $e$ & $\omega[deg]$ & $i[deg]$ & $M/M_\odot$  & $R/R_\odot$ & $K_{\rm tide}[m/s]^{\rm a} $ & $e K_{\rm orb}[m/s] $ \\
\hline
WASP-19 b                         &       1.17   &       0.79   &     0.0046   &       3.00   &      79.40   &       0.97   &       0.99   &       2.83   &       1.19  \\
WASP-43 b                         &       1.78   &       0.81   &              &              &      82.60   &       0.58   &       0.93   &       8.90   &             \\
WASP-18 b                         &      10.11   &       0.94   &     0.0085   &     -92.10   &      86.63   &       1.24   &       1.36   &      31.94   &      15.38  \\
WASP-12 b                         &       1.40   &       1.09   &              &              &      86.00   &       1.35   &       1.60   &       4.78   &             \\
OGLE-TR-56 b                      &       1.30   &       1.21   &              &              &      78.80   &       1.17   &       1.32   &       2.12   &             \\
HAT-P-23 b                        &       2.09   &       1.21   &     0.1060   &     118.00   &      85.10   &       1.13   &       1.20   &       3.61   &      39.03  \\
WASP-33 b                         &       4.59   &       1.22   &     0.174$^{\rm c}$ & -89.0$^{\rm c}$ &      87.67   &       1.50   &       1.44   &       5.89   & 120.0            \\
HD 41004 B b                      &      18.40   &       1.33   &     0.0810   &     178.50   & no transit   &       0.40   &       0.48$^{\rm b}$   &       3.61   &     517.48  \\
WASP-4 b                          &       1.12   &       1.34   &              &              &      88.80   &       0.93   &       1.15   &       1.13   &              \\
CoRoT-14 b                        &       7.60   &       1.51   &              &              &      79.60   &       1.13   &       1.21   &       4.11   &              \\
SWEEPS-11                         &       9.70   &       1.80   &              &              &      84.00   &       1.10   &       1.45   &       7.04   &              \\
HAT-P-7 b                         &       1.80   &       2.20   &              &              &      84.10   &       1.47   &       1.84   &       1.02   &              \\
WASP-14 b                         &       7.72   &       2.24   &     0.0903   &     254.90   &      84.79   &       1.32   &       1.30   &       1.55   &      90.59  \\
OGLE2-TR-L9 b                     &       4.34   &       2.49   &              &              &      79.80   &       1.52   &       1.53   &       1.85   &             \\
XO-3 b                            &      11.79   &       3.19   &     0.2600   &     345.80   &      84.20   &       1.21   &       1.38   &       3.07   &     384.77  \\
HAT-P-2 b                         &       8.74   &       5.63   &     0.5171   &     185.22   &      90.00   &       1.36   &       1.64   &       5.04   &     482.55  \\
HIP 13044 b                       &       1.25   &      16.20   &     0.2500   &     219.00   & no transit   &       0.80   &       6.70   &       2.74   &      30.01  \\
\hline
\end{tabular}
\\
$^{\rm a}$ Semi-amplitude ((max-min)/2) found by evaluating eq.\ref{eq:vtide} over the orbit. \\
$^{\rm b}$ Radius of HD 41004 B b  not listed in exoplanets.eu, so we use the main sequence value of $0.48\ R_\odot$. \\
$^{\rm c}$ The values of $e$ and $\omega$ for WASP-33 b were found by \citet{2011MNRAS.tmp.1075S}, but rejected as unphysical. Their main results instead fix $e=0$. \\
\end{table*}

%% file: ms_mnras.bbl
\begin{thebibliography}{}
\bibitem[Bedding \& Kjeldsen(2007)]{2007CoAst.150..106B} Bedding, T.~R., \& Kjeldsen, H.\ 2007, Communications in Asteroseismology, 150, 106
\bibitem[Butler et al.(1996)]{1996PASP..108..500B} Butler, R.~P., Marcy, G.~W., Williams, E., McCarthy, C., Dosanjh, P., \& Vogt, S.~S.\ 1996, \pasp, 108, 500
\bibitem[Dziembowski(1977)]{1977AcA....27..203D} Dziembowski, W.\ 1977, {\it Acta Astronomica}, 27, 203
\bibitem[Goldreich \& Nicholson(1989)]{1989ApJ...342.1079G} Goldreich, P., \& Nicholson, P.~D.\ 1989, \apj, 342, 1079
\bibitem[Laughlin \& Chambers(2001)]{2001ApJ...551L.109L} Laughlin, G., \& Chambers, J.~E.\ 2001, \apjl, 551, L109
\bibitem[Loeb \& Gaudi(2003)]{2003ApJ...588L.117L} Loeb, A., \& Gaudi, B.~S.\ 2003, \apjl, 588, L117
\bibitem[Lovis et al.(2006)]{2006SPIE.6269E..23L} Lovis, C., et al.\ 2006,  \procspie, 6269,
\bibitem[Murray \& Correia(2010)]{2010exop.book...15M} Murray, C.~D., \& Correia, A.~C.~M.\ 2010, Exoplanets, 15
\bibitem[Murray \& Dermott(1999)]{1999ssd..book.....M} Murray, C.~D., \& Dermott, S.~F.\ 1999, Solar system dynamics by Murray, C.~D., 1999,
\bibitem[Pfahl et al.(2008)]{2008ApJ...679..783P} Pfahl, E., Arras, P., \& Paxton, B.\ 2008, \apj, 679, 783
\bibitem[Sirko \& Paczy{\'n}ski(2003)]{2003ApJ...592.1217S} Sirko, E., \& Paczy{\'n}ski, B.\ 2003, \apj, 592, 1217
\bibitem[Smith et al.(2011)]{2011MNRAS.tmp.1075S} Smith, A.~M.~S., Anderson, D.~R., Skillen, I., Collier Cameron, A., \& Smalley, B.\ 2011, \mnras, 1075 
\bibitem[Terquem et al.(1998)]{1998ApJ...502..788T} Terquem, C., Papaloizou, J.~C.~B., Nelson, R.~P., \& Lin, D.~N.~C.\ 1998, \apj, 502, 788 
\bibitem[Triaud et al.(2010)]{2010A&A...524A..25T} Triaud, A.~H.~M.~J., et al.\ 2010, \aap, 524, A25
\bibitem[Unno et al.(1989)]{1989nos..book.....U} Unno, W., Osaki, Y., Ando, H., Saio, H., \& Shibahashi, H.\ 1989, Nonradial oscillations of stars, Tokyo: University of Tokyo Press, 1989, 2nd ed.,
\bibitem[Welsh et al.(2010)]{2010ApJ...713L.145W} Welsh, W.~F., Orosz, J.~A., Seager, S., Fortney, J.~J., Jenkins, J., Rowe, J.~F., Koch, D., \& Borucki, W.~J.\ 2010, \apjl, 713, L145
\bibitem[Willems \& Aerts(2002)]{2002A&A...384..441W} Willems, B., \& Aerts, C.\ 2002, \aap, 384, 441
\bibitem[Winn(2011)]{2011EPJWC..1105002W} Winn, J.~N.\ 2011, Detection and Dynamics of Transiting Exoplanets, St.~Michel l'Observatoire, France, Edited by F.~Bouchy; R.~D{\'{\i}}az; C.~Moutou; EPJ Web of Conferences, Volume 11, id.05002, 11, 5002
\bibitem[Wright(2005)]{2005PASP..117..657W} Wright, J.~T.\ 2005, \pasp,  117, 657
\end{thebibliography}
